\documentstyle[12pt]{article}
\newcommand{\bea}{\begin{eqnarray}}
\newcommand{\eea}{\end{eqnarray}}
\newcommand{\om}{\omega}

\begin{document}

\date{}
\title{Of Some Theoretical Significance: Implications of Casimir Effects}
\author{G. Jordan Maclay \\
%EndAName
{\small {Quantum Fields LLC}}\\
{\small {20876 Wildflower Lane}}\\
{\small {Richland Center, Wisconsin 53581}}\\
\\
and\\
\\
Heidi Fearn\footnote{Permanent address: Department of Physics, California
State University, Fullerton, California 92834} \ and Peter W. Milonni\\
{\small {Theoretical Division (T-4)}}\\
{\small {Los Alamos National Laboratory}}\\
{\small {Los Alamos, New Mexico 87545}}}
\maketitle

\begin{abstract}
In his autobiography Casimir barely mentioned the Casimir effect, but
remarked that it is ``of some theoretical significance" \cite{casbook}. We
will describe some aspects of Casimir effects that appear to be of
particular significance now, more than half a century after Casimir's famous
paper \cite{cas1}.
\end{abstract}

\section{Introduction}

Let us first recall that Casimir discovered his effect as a byproduct of
some applied industrial research in the stability of colloidal suspensions
used to deposit films in the manufacture of lamps and cathode ray tubes.  In
the 1940s J.T.G. Overbeek at the Philips Laboratory studied the properties
of suspensions of quartz powder, and experiments indicated that the theory
of colloidal stability he had developed with E.J.W. Verwey could not be
entirely correct. Better agreement between theory and experiment could be
obtained if the interparticle interaction somehow fell off more rapidly at
large distances than had been supposed. Overbeek suggested that this might
be related to the finite speed of light, and his suggestion prompted his
co-workers Casimir and Polder to reconsider the theory of the van der Waals
interaction with retardation included. They concluded that Overbeek was
right: retardation causes the interaction to vary as $r^{-7}$ rather than $%
r^{-6}$ at large intermolecular separations $r$.

Intrigued by the simplicity of the result, Casimir sought a deeper
understanding. A conversation with Bohr led him to an interpretation in
terms of zero-point energy. Then he went further with the idea of zero-point
energy and showed that two perfectly conducting parallel plates should be
attracted to each other as a consequence of the change they create in
zero-point field energy:

\begin{quotation}
{\small Summer or autumn 1947 (but I am not absolutely certain that it [was]
not somewhat earlier or later) I mentioned my results to Niels Bohr, during
a walk. That is nice, he said, that is something new. I told him that I was
puzzled by the extremely simple form of the expresssions for the interaction
at very large distance and he mumbled something about zero-point energy.
That was all, but it put me on a new track. }

{\small I found that calculating changes of zero-point energy really leads
to the same results as the calculations of Polder and myself ... }

{\small On May 29, 1948 I presented my paper ``On the attraction between two
perfectly conducting plates" at a meeting of the Royal Netherlands Academy
of Arts and Sciences. It was published in the course of the year ... \cite%
{cas2} }
\end{quotation}

At about the same time the observation of the Lamb shift led to the
interpretation of that effect in terms of changes in zero-point energy, or
vacuum fluctuations, but Casimir's thinking was independent of this
development:

\begin{quotation}
{\small ... I was not at all familiar with [that work]. I went my own,
somewhat clumsy way ... I do not think there were outside influences ... %
\cite{cas2}. }
\end{quotation}

The Casimir force between conducting plates is a more palpable consequence
of zero-point field than, for instance, the Lamb shift. \ It is perhaps for
this reason that it now appears to be the most widely cited example of how
vacuum fields and their fluctuations can have observable manifestations. \
The current interest owes much to recent experiments that unambiguously
confirm Casimir's prediction and allow the experimental investigation of
such things as finite conductivity and temperature corrections to the
Casimir force between plates \cite{lamor}, \cite{moh}.

The experimental verification of Casimir's prediction is often cited as
proof of the reality of the vacuum energy density of quantum field theory. \
Yet, as Casimir himself observed, other interpretations are possible:

\begin{quotation}
The action of this force [between parallel plates] has been shown by clever
experiments and I\ think we can claim the existence of the electromagnetic
zero-point energy without a doubt. But one can also take a more modest
point of view. \ Inside a metal there are forces of cohesion and if you take
two metal plates and press them together these forces of cohesion begin to
act. \ On the other hand you can start with one piece and split it. \ Then
you have first to break chemical bonds and next to overcome van der Waals
forces of classical type and if you separate the two pieces even further
there remains a curious little tail. The Casimir force, {\it sit venia verbo},
is the last but also the most elegant trace of cohesion energy \cite{cas3}.
\end{quotation}
Casimir effects have also been derived and interpreted in terms of {\it source}
fields in both conventional \cite{pwm} and nonconventional \cite{schw}
quantum electrodynamics.

Casimir effects result from changes in the ground-state fluctuations of a
quantized field that occur due to the boundary conditions. Casimir effects
occur for all quantum fields and can also arise from the choice of topology.
In the special case of the vacuum electromagnetic field with dielectric
or conductive boundaries, various approaches suggest that Casimir forces 
can be regarded as macroscopic manisfestations of many-body
retarded van der Waals forces \cite{pwm}, \cite{Power}.

Zero-point field energy density is a simple and inexorable consequence of 
quantum theory, but it brings puzzling inconsistencies with
another well verified theory, general relativity. \ The total energy density
of the vacuum would be expected to provide a cosmological constant of the
type introduced by Einstein in order to have static solutions of his field
equations. \ The predicted electromagnetic quantum vacuum energy density is
enormous (about $10^{114}$ J/m$^3$ or, in terms of mass,
$10^{95}$ g/cm$^3$ if the Planck length of 10$^{-35}$ m is used to provide a
cut-off), and for an infinite flat universe would imply an outward
zero-point pressure that would rip the 
universe apart \cite{Vis}. Astronomical data, on the other hand,
indicate that any such
cosmological constant must be $\sim 4$ eV/mm$^3$, or $10^{-29}$ g/cm$^3$
when expressed as mass \cite{Ost}. The discrepancy between theory and
observation is about 120 orders of magnitude, arguably the greatest
quantitative discrepancy between theory and observation in the history of
science \cite{weinberg}, \cite{adler}! There are numerous approaches to
solve this ``cosmological constant problem,'' such as renormalization,
supersymmetry, string theory, and quintessence, but as yet this remains 
an unsolved problem.

The Casimir effect is important as well in connection with other aspects of
cosmology and space-time physics. Fluctuations in vacuum field energy are
believed by cosmologists to be responsible for the origin of the universe.
These fluctuations may have provided the primordial irregularities required
to form stars and galaxies, and may be the source of the cosmic temperature
fluctuations uncovered by the COBE satellite in 1992.

The possibility of a ``traversable wormhole'' tunnel in space-time \cite%
{morris} is attributable to the modification of the vacuum by the Casimir
effect, and in particular to the negative energy density between the caps of
the wormhole.  There is an interesting question, however, about whether 
the positive energy density associated with the caps will result in a 
net energy density that is insufficiently negative (Visser 1996, pp 121-6).
 
Observable consequences of focusing vacuum fluctuations with a
parabolic reflector have recently been  predicted \cite{Ford}. 
Casimir effects can also arise
from dynamical constraints, such as moving mirrors or varying
gravitational fields, that alter the vacuum. A sudden displacement of a
reflecting boundary, for instance, is not communicated to a point at a
distance $d$ from the boundary until a time $d/c$, and a consequence of this
is that radiation is generated, i.e., particles are created. The effect is
very weak unless enormous accelerations are imagined. However, if one of the
plates in the original Casimir example is made to oscillate resonantly with
the photon propagation time in the cavity, there is an amplification of the
effect that might make the radiation observable \cite{reynaud}. Many of
the recent predictions of vacuum-field effects are, to say the least,
not readily observable
\cite{Scharn}, \cite{Milonni meas}. The significance of Casimir's work in
this context is that it makes an experimentally verifiable prediction based
on the quantum vacuum, and thereby lends support to these other predictions
that rely on quantum vacuum theory.

Another vacuum effect that has received much attention is the Unruh-Davies
effect: a detector (or atom) moving with uniform acceleration in the vacuum
responds as if it is at rest in a thermal field of temperature $T=\hbar
a/2\pi kc$, where $a$ is the proper acceleration and $k$ is Boltzmann's
constant. Vacuum fluctuations are in
effect promoted to thermal fluctuations. Unfortunately the accelerations
required for one to seriously contemplate an experimental observation of the
effect are prohibitively large. (A temperature 1 pK corresponds to an
acceleration of $2.5 \times 10^8$ m/sec$^2$.)

\section{Vacuum Friction}

Consider instead the case of an atom moving in an isotropic {\it thermal}
field. In this case there is an effect that depends on the atom's {\it 
velocity}: an atom with velocity $v$ experiences a drag force 
\begin{equation}
F=-\left( {\frac{\hbar \omega }{c^{2}}}\right) (p_{1}-p_{2})B_{12}\left(
\rho (\omega )-{\frac{\omega }{3}}{\frac{d\rho }{d\omega }}\right) v\ ,
\label{eq1}
\end{equation}%
where $\omega $ is the transition frequency between the lower state 1 and
the upper state 2, $p_{1}$, $p_{2}$ are the state occupation probabilities, 
$B_{12}$ is the Einstein $B$ coefficient for absorption, and $\rho(\omega )$
is the spectral energy density of the field. (For simplicity we restrict
ourselves to two nondegenerate energy levels of the atom.) This result was
obtained by Einstein \cite{ein}, who showed that the increase in the atoms'
kinetic energy upon absorption and emission of radiation is balanced on
average by the drag force if the equilibrium $\rho(\omega )$ is the Planck
spectrum.

What does this have to do with Casimir effects or the vacuum? Let us
note first that, for the vacuum spectral energy density $\rho_{0}(\omega)
=\hbar\omega^{3}/2\pi^{2}c^{3}$, the drag force vanishes. This is as it
should be: Lorentz invariance of the vacuum does not allow a
velocity-dependent force. But what happens if we arrange for the
zero-temperature spectral density of the electromagnetic field to be
different from the $\rho_{0}(\omega)$ of infinite free space? 

One way to obtain a zero-temperature spectral density different from
$\rho_0(\om)$, of course, is to introduce conducting surfaces.
Local changes in mode density and therefore vacuum energy
density are induced by the presence of curved surfaces, and, depending on
whether the curvature is positive or negative, the force between the surface
and the particle may be repulsive or attractive \cite{Deutsch}. Indeed,
whenever there is an inhomogeneous vacuum energy density, there will a net
force on a polarizable neutral particle given by $\frac{1}{2}\alpha 
\overrightarrow{\nabla }\langle E(x)^{2}\rangle .$
The simplest example of using a surface to alter vacuum modes 
is a perfectly conducting, infinite wall. The change in the 
vacuum field energy due to the wall produces in this case the well-known
Casimir-Polder interaction: for sufficiently large distances $d$ from the
wall this interaction is $V(d)=-3\alpha\hbar c/8\pi d^4$, where $\alpha$ 
is the static polarizability of the (ground-state) atom. This effect
has been accurately verified in the elegant experiments of Sukenik 
{\it et al} \cite{suk}.

Now let the atom move parallel to the wall with velocity $v$. In this
case, provided the wall is not an idealized perfect conductor, there
is a velocity-dependent force $F(v)$ acting along the direction of motion of the
atom. This force can be associated physically with the effect of the finite
conductivity of the wall material on the image field of the atom. The
functional form of $F(v)$ depends sensitively on how the dielectric
function of the wall material varies with frequency \cite{barton1}. Pendry
\cite{pendry} has discussed the possibility of a frictional force
when two infinite parallel mirrors separated by a fixed distance are in relative
motion, and finds ``large frictional effects comparable to everyday
frictional forces provided that the materials have resistivities of the
order of 1 m$\Omega$ and that the surfaces are in close proximity."
As in the case of an atom moving with respect to a wall, the form of
the frictional force depends sensitively on the dielectric function.
In fact, a {\it Gedanken} experiment suggests that lateral Casimir forces are
present even for ideal finite conducting parallel planes, otherwise it would
be possible to construct a device that would extract a net positive energy
from the vacuum in each cycle of its operation \cite{maclay}.

\section{Technological Implications}
We have already alluded to what may be some profoundly important aspects of
the quantum vacuum, and have noted that the reality of various Casimir
effects lends credibility to predictions of various vacuum field
effects that lie fantastically beyond the pale of experiment. Of course 
Casimir effects are also of interest in their own right and, 
if anything, this interest appears to be growing. Moreover, recent 
work -- including that on vacuum friction -- suggests that Casimir 
effects may be of some practical significance.

Casimir effects will be significant in microelectromechanical 
systems (MEMS) if (when) further miniaturization is realized
\cite{serry}. Smaller distances between MEMS components are desirable
in electrostatic actuation schemes because they permit
smaller voltages to be used to generate larger forces and torques. 
MEMS currently employed in sensor and actuator technology have component 
separations on the order of microns, where Casimir effects are negligible. 
However, the Casimir force per unit area between perfectly conducting plates,
$F=-\pi^2\hbar c/240d^4$, increases rapidly as the separation $d$ is
decreased; at a separation of 10 nm, $F\sim 1$ atm. 

Serry {\it et al} \cite{serry} have considered an idealized MEMS component
resembling the original Casimir example of two parallel plates, except 
that one of the plates is connected to a stationary surface by a linear 
restoring force and can move along the direction normal to the plate surfaces. 
The Casimir force between the two plates, together with the restoring
force acting on the moveable plate, results in an ``anharmonic Casimir
oscillator" exhibiting bistable behavior as a function of the plate separation.
This suggests the possibility of a switching mechanism, based on the
Casimir effect, that might be used in the design of sensors and deflection
detectors \cite{serry}. The analysis also suggests that the Casimir effect
might be responsible in part for the ``stiction" phenomenon in which
micromachined membranes are found to latch onto nearby surfaces. 

An experimental demonstration of the Casimir effect in a nanometer-scale
MEMS system has recently been reported \cite{chan}. In the experiment
the Casimir attraction between a 500 $\mu$m-square plate 
suspended by torsional rods and a gold-coated sphere of 
radius 100 $\mu$m was observed as a sharp
increase in the tilt angle of the plate as the sphere-plate separation
is reduced from 300 nm to 75.7 nm. This ``quantum mechanical actuation"
of the plate suggests ``new possibilities for novel actuation schemes in
MEMS based on the Casimir force" \cite{chan}. 

\section{Complications and Approximations}

Calculations of Casimir forces for situations more complicated than two
parallel plates are notoriously difficult, and one has little intuition
even as to whether the force should be attractive or repulsive for any
given geometry. The fact that the Casimir force on a perfectly conducting
spherical shell is {\it repulsive}, as first discovered by Boyer
\cite{boyer1}, surprised even Casimir, who presumed that the force would
be attractive \cite{cas4}. Since Boyer's work a fairly large literature
has grown around problems of calculating Casimir forces for perfectly
conducting spheres, cubes, cylinders, wedges, and other geometries.

In the case of dielectrics the situation is even more complicated, as
can be appreciated from the Lifshitz theory for the ``simple"
example of two parallel walls (see, for instance, Milonni 1994, pp 219-33). 
Computation of the numerical value of the force per unit area requires a
knowledge of the complex refractive index as a function of frequency as
well as the deviation from perfect surface smoothness; therefore, as 
discussed by Lamoreaux \cite{lamor2} and Klimchitskaya {\it et al} 
\cite{klim}, for instance, accurate computations 
require auxiliary measurements of various properties of the surfaces. 

It would be very useful to have approximate methods for the calculation
of Casimir forces for arbitrarily shaped bodies. 
The obvious and simplest approximation is to add up {\it pairwise}
van der Waals forces \cite{milonniandshih}. 
Consider, for instance, an atom $A$ at a distance
$d$ from a half-space of $N$ atoms per unit volume, and suppose that
all the atoms are identical and that $d$ is large enough that the
interaction between $A$ and each atom of the ``wall" is the 
retarded van der Waals interaction $V(r)=-23\hbar c\alpha^2/4\pi r^7$. 
Adding the pairwise interactions between $A$ and all the wall atoms,
one easily finds
\begin{equation}
V(d)=-{23\over 40}{\hbar c\alpha\over d^4}N\alpha \ .
\label{eq2}
\end{equation}
Now if we use the Clausius-Mossotti relation between 
$N\alpha$ and the dielectric constant $\epsilon$, and assume that 
the limit $\epsilon\rightarrow\infty$ should correspond to a perfectly
conducting wall, then the potential energy of $A$ when it is at a
distance $d$ from a perfectly conducting plate should be
\begin{equation}
V(d)=-{69\over 160\pi}{\alpha\hbar c\over d^4} 
\label{eq3}
\end{equation}
in the pairwise approximation. This is 15\% larger than the
Casimir-Polder result cited earlier. A similar calculation of the 
pairwise van der Waals force per unit area between two parallel walls gives
\begin{equation}
F(d)=-{207\hbar c\over 640\pi^2d^4} \ ,
\label{eq4}
\end{equation}
which is 20\% smaller than the Casimir result \cite{cas1}.

In light of the stark simplicity of the pairwise approach, these results
are certainly encouraging. However, these two examples pretty much
exhaust the supply of known, exact results for the Casimir interaction
of disconnected objects. Let us consider therefore some known results 
for connected objects.

We have already alluded to the Casimir energy for a perfectly conducting
spherical shell. A calculation of the pairwise van der Waals energy of a
spherical ball of radius $a$ gives $V(a)=(207\hbar c/1536\pi a)[(\epsilon-1)/
(\epsilon+2)]^2$ if we ignore dispersion and assume again that $N\alpha$ 
and $\epsilon$ are related by the Clausius-Mossotti formula. Thus
$\lim_{\epsilon\rightarrow\infty}V(a)=.043\hbar c/a$ for a spherical ball
is of the same
order as the exact result ($.09\hbar c/a$) for the conducting spherical
shell \cite{boyer1} and gives the correct, ``counter-intuitive" sign.
(Note: We disregard infinities associated with the divergence of the
van der Waals interaction when the intermolecular spacing goes to zero,
as assumed in our continuum model. That is, we retain only what
Barton \cite{barton2} refers to as the ``pure Casimir term.") 

Unfortunately the surprising degree of accuracy of this (relatively)
simple approach in these examples seems fortuitous. For the case 
of an infinitely long conducting cylindrical shell of radius $a$, 
the pairwise approach gives a Casimir energy of zero, whereas 
several authors have found that
there is an attractive force. (See Barton 2000, Reference \cite{barton2},
and references therein.)
In the case of a conducting cube of side $a$ the exact calculation yields
$V(a)=.092/a$, whereas Barton finds that the pairwise approximation gives
an {\it attractive} Casimir term \cite{barton2}. (After initially 
obtaining a repulsive force, and checking the laborious calculations 
after learning of Barton's result, we have confirmed that the ``pure
Casimir term" is indeed attractive in the pairwise approximation.) 

Ambjorn and Wolfram \cite{amb} remarked that ``[the pairwise approximation]
... in the case of two parallel planes ... leads to a correct Casimir
energy," but that ``this result is probably fortuitous." They support 
this claim by remarking that, ``according to [the pairwise approximation] 
the Casimir forces between conducting surfaces would ... always be attractive,"  
whereas for the cube, for instance, the actual Casimir force is repulsive,
as they showed. We note, however, that the pairwise approximation for the
sphere gives in fact a repulsive force for the pure Casimir term.

It appears then that there is still no reliable 
approximation to the evaluation of Casimir forces for arbitrarily 
shaped bodies. It is worth noting, however, that Schaden and Spruch 
\cite{schaden} have developed a semiclassical approach that might 
lend itself to workable approximations for arbitrary geometries.

\section{Acknowledgement}

In this tribute to Casimir we have tried to convey our strong belief
that, after all these years, there is still much to appreciate and
learn about Casimir effects. We are grateful not only to Casimir but
to our many colleagues whose work has kept the subject alive and well.
For the discussion in the final section we are particularly grateful
to Gabriel Barton for sharing his results with us prior to their
publication. GJM would like to thank the NASA Breakthrough Propulsion
Program for its support of this work.

\end{document}